\documentclass[11pt, a4paper]{article}
\usepackage{graphicx}
\usepackage{amsmath}
\usepackage{amssymb}
\usepackage{geometry}
\usepackage{cite}
\usepackage{xcolor}

\title{Refined Sensitivity Estimates for Single-Molecule Magnet Dark Matter Detectors}

\author{
    Andrew Eberhardt$^{1,2}$, Tomoya Fukui$^{3,4}$, Ryosuke Takehara$^{3,4}$, Ryotaro Ohno$^5$, \\ 
    Yuta Mizukami$^5$, Kenichiro Hashimoto$^6$, Takanori Fukushima$^{3,4}$,  \\
    Shigeki Matsumoto$^1$, Tom Melia$^{1,2}$, Kouki Nozaki$^{1,2,7}$, Surjeet Rajendran$^8$ \\
    \small $^1$Kavli IPMU (WPI), UTIAS, The University of Tokyo, Kashiwa, Chiba 277-8583, Japan \\
    \small$^2$Center for Data-Driven Discovery, Kavli IPMU (WPI), UTIAS, The University of Tokyo, \\ 
    \small Kashiwa, Chiba 277-8583, Japan \\
    \small $^3$ Laboratory for Chemistry and Life Science, 
     Institute of Integrated Research, \\ \small Institute of Science Tokyo, 4259 Nagatsuta, Midori-ku, Yokohama, 226-8501, Japan \\
    \small $^4$	Research Center for Autonomous Systems Materialogy (ASMat), Institute of Integrated Research, \\ \small Institute of Science Tokyo, 4259 Nagatsuta, Midori-ku, Yokohama, 226-8501, Japan\\    
    \small $^5$Department of Physics, Tohoku University, Sendai, 980-8578, Japan \\
    \small $^6$Department of Physics, Kyoto University, Kyoto 606-8502, Japan \\
    \small $^7$Department of Physics, Graduate School of Science, The University of Tokyo, Tokyo 113-0033,
Japan \\
    \small $^8$The Department of Physics and Astronomy, The Johns Hopkins University, Baltimore, MD 21218 
}

\date{\today}

\begin{document}

\maketitle

\begin{abstract}
We revisit the sensitivity of Single Molecule Magnet (SMM) crystals as detectors for low-mass dark matter. In previous work, we established the concept of the ``magnetic bubble chamber,'' where energy deposited by dark matter triggers a magnetic avalanche in a metastable crystal. The original sensitivity estimates relied on a conservative criterion requiring the spin relaxation time to be strictly shorter than the thermal diffusion time. Here, we demonstrate that this criterion effectively ignores the stochastic nature of spin relaxation. We derive a refined analytic estimate which accounts for the fraction of spins that relax even when diffusion is fast. We show that the Zeeman energy released by this fraction contributes to local heating, significantly lowering the energy threshold for avalanche formation. We present simulation results confirming this effect and report on experimental verification of the assumed low-temperature thermal properties of two representative SMM crystals, Mn$_{12}$-acetate and Mn$_{32}$. Together, these efforts extend this pathfinder program toward the realization of SMM-based detectors with controlled material properties and enhanced dark matter sensitivity.
\end{abstract}

\section{Introduction}

The search for dark matter in the sub-GeV mass range requires novel sensor technologies capable of detecting meV-scale energy depositions. The sub-GeV dark matter paradigm has attracted considerable experimental attention over the past decade, motivated by the possibility that dark matter couples to electrons or nuclei at sub-keV recoil energies~\cite{Essig:2012,Essig:2016,Graham:2012}. A range of experiments, including SENSEI~\cite{SENSEI:2020dpa}, SuperCDMS~\cite{SuperCDMS:2018mne}, CRESST-III~\cite{CRESST:2016qpj}, EDELWEISS~\cite{Arnaud:2020}, and DAMIC~\cite{DAMIC:2021crr}, have pushed detection thresholds toward the sub-hundred-eV regime. A striking observation in this context is the appearance of sharply rising event rates at low energies across many of these experiments, well in excess of known background estimates; this phenomenon has been systematically catalogued by the EXCESS collaboration~\cite{Fuss:2022fxe}.

In a previous proposal~\cite{Bunting:2017}, we introduced the use of Single Molecule Magnets (SMMs) as ``magnetic bubble chambers.'' The SMM detector concept is designed to address precisely this challenge: by exploiting the macroscopic stored magnetic energy of a metastable molecular crystal, energy deposits far below the eV scale can trigger a macroscopic, readable avalanche signal throughout the bulk of the crystal. An SMM crystal is prepared in a metastable magnetic state. A localized energy deposition---such as that from a dark matter scattering event---heats a small region, reducing the magnetic relaxation time of the molecular spins. If these spins relax to the ground state, they release stored Zeeman energy, which generates further heating and triggers a self-sustaining magnetic avalanche.

An important distinguishing feature of the SMM detector architecture is that it operates as a true bulk detector, in the spirit of a bubble chamber: the signal originates and propagates within the crystal volume itself, rather than being sensed by electrodes or phonon collectors at the crystal surface. This stands in contrast to the semiconductor and bolometric detectors that contribute to the EXCESS observations~\cite{Fuss:2022fxe}, where surface-related effects---such as stress-induced phonon emission, charge trapping, and athermal phonon leakage at interfaces---are among the leading candidate explanations for the anomalous low-energy event rates. An SMM bulk detector is expected to have a qualitatively different background profile, and the development of the SMM platform therefore also offers a complementary handle on the origin of the low-energy excesses observed in current experiments.

Since the initial proposal, the mechanism of particle-induced magnetic avalanches in SMMs has been experimentally demonstrated. Chen \textit{et al.}~\cite{Chen:2020} reported the first observation of magnetic avalanches triggered by $\alpha$-particle scattering in Mn$_{12}$-acetate, with a composition of [Mn$_{12}$(CH$_3$CO$_2$)$_{16}$(H$_2$O)$_4$O$_{12}$]$\cdot$2(CH$_3$CO$_2$H)$\cdot$4(H$_2$O), hereafter denoted Mn$_{12}$ac~\cite{Lis:1980}. Subsequent work by Kohn \textit{et al.}~\cite{Kohn:2025} confirmed these findings at lower temperatures (800~mK) and characterized the avalanche dynamics in greater detail. While these experiments operated at MeV energy thresholds due to the specific materials and conditions used, they validate the fundamental detection principle.

In this work, we refine the theoretical framework used to estimate the ignition threshold for particle-induced avalanches. The original sensitivity projections in Ref.~\cite{Bunting:2017} implemented a conservative ``hard'' criterion: an avalanche was assumed to occur only when the characteristic spin-relaxation time is shorter than the thermal diffusion time out of the heated region. Here we show that this criterion is overly restrictive because it treats spin relaxation as deterministic rather than stochastic. Even when diffusion is fast, a nonzero fraction of spins will relax during the diffusion window, releasing Zeeman energy that contributes to local heating. Accounting for this integrated energy release yields a parametrically improved ignition condition and lowers the predicted threshold for avalanche formation.

We validate this framework with 3D simulations of coupled thermal and spin dynamics and with new low-temperature measurements of the specific heat and thermal diffusivity of representative SMM crystals. These measurements directly test the assumptions underlying the analytic treatment---most notably the Debye-like $c(T)\propto T^3$ behavior that governs the temperature response to localized energy deposition---and provide experimentally grounded inputs for realistic sensitivity projections. The remainder of this paper is organised as follows. Section~2 derives the refined analytic ignition criterion. Section~3 describes the 3D simulations that validate it. Section~4 presents the low-temperature calorimetric and thermal-transport measurements of Mn$_{12}$ac and a crystalline sample of [Mn$_{32}$O$_8$(OH)$_6$(Me-sao)$_{14}$(CH$_3$CO$_2$)$_{18}$Br$_8$(H$_2$O)$_{10}$](OH)$_2$ (hereafter Mn$_{32}$, see Ref.~\cite{manoli:2011}). Section~5 concludes.

\section{Refined Avalanche Dynamics}

In Ref.~\cite{Bunting:2017}, the criterion for triggering an avalanche was derived by comparing timescales in a heated volume of radius $R$: the thermal diffusion time $t_d$ and the magnetic relaxation time $t_f(T)$. An avalanche was deemed possible only when $t_f < t_d$, i.e.\ spins relax before heat escapes. Here we refine this by considering the full energy balance, showing that the relevant condition is more subtle and considerably more favourable.

When energy $E_{\text{dep}}$ is deposited, it heats a spherical volume $V = \frac{4}{3}\pi R^3$. The resulting local temperature $T$ is set by the Debye phonon specific heat $c(T) = \beta T^3$:
\begin{equation}
    E_{\text{dep}} = V \int_0^T c(T') \, dT' = \frac{\pi R^3 \beta T^4}{3}.
\label{eq:edep}
\end{equation}
The heat remains within this volume for a characteristic diffusion time
\begin{equation}
    t_d \approx \frac{R^2}{\alpha},
\label{eq:td}
\end{equation}
where $\alpha$ is the thermal diffusivity. The magnetic relaxation time follows an Arrhenius law,
\begin{equation}
    t_f(T) = \tau_0 \exp\!\left(\frac{\tilde{U}(B)}{k_B T}\right),
\label{eq:tf}
\end{equation}
where $\tilde{U}(B) = U_{\rm eff} - 2\mu_B J B$ is the effective barrier, reduced from the zero-field value $U_{\rm eff}$ by the Zeeman energy $2\mu_B J B$ in an applied field $B$.

\subsection*{Stochastic relaxation in the fast-diffusion regime}

We focus on the regime $t_d < t_f$, which is precisely where the original hard criterion of Ref.~\cite{Bunting:2017} predicts \emph{no} avalanche. In this regime diffusion is fast, but spin relaxation is a stochastic process: even though the \emph{mean} relaxation time exceeds the diffusion window, a nonzero fraction of spins will relax during $t_d$. For $t_d \ll t_f$, this fraction is well approximated by the leading-order expansion of the Poisson survival probability,
\begin{equation}
    f = 1 - e^{-t_d/t_f(T)} \approx \frac{t_d}{t_f(T)},
\label{eq:fraction}
\end{equation}
where the linearisation is valid precisely in the regime of interest $t_d/t_f \ll 1$. Note that when $t_d > t_f$ the fraction $f \to 1$ and essentially all spins relax --- the system is firmly in the avalanche regime of Ref.~\cite{Bunting:2017} and no refined treatment is needed. The interesting new physics arises in the $t_d < t_f$ window, where $f$ is small but nonzero.

Each relaxing spin releases Zeeman energy $\Delta E_{\rm Zee} \equiv 2J\mu_B B$, deposited locally within the heated volume. The total Zeeman energy released by the fraction $f$ of the $\rho_s V$ spins in the volume is
\begin{equation}
    E_{\text{rel}} = \rho_s V \,\Delta E_{\rm Zee}\, f.
\label{eq:erel_def}
\end{equation}
Substituting Eqs.~(\ref{eq:td})--(\ref{eq:fraction}) and $V = \frac{4}{3}\pi R^3$:
\begin{equation}
    E_{\text{rel}} = \frac{4\pi \rho_s \Delta E_{\rm Zee}}{3\,\alpha\, t_f(T)}\, R^5.
\label{eq:erel}
\end{equation}
The key observation is the $R^5$ scaling of $E_{\rm rel}$, which contrasts with the $R^3$ scaling of $E_{\rm dep}$ in Eq.~(\ref{eq:edep}). For a sufficiently large initial heated radius, the released energy will therefore always become comparable to the deposited energy, regardless of how small $f$ is.

\subsection*{The critical radius and energy threshold}

Avalanche ignition requires positive feedback: the Zeeman energy released during $t_d$ must be sufficient to sustain (and amplify) the local temperature gain against diffusive losses. The order-unity condition for this feedback to develop is
\begin{equation}
    E_{\rm rel} \sim E_{\rm dep}.
\label{eq:ignition}
\end{equation}
Substituting Eqs.~(\ref{eq:edep}) and (\ref{eq:erel}) into Eq.~(\ref{eq:ignition}) gives a condition on the heated radius at temperature $T$:
\begin{equation}
    R_c^2(T) = \frac{\alpha\,\tau_0\,\beta\,T^4}{4\,\rho_s\,\Delta E_{\rm Zee}}\,
    \exp\!\left(\frac{\tilde{U}(B)}{k_B T}\right).
\label{eq:rc}
\end{equation}
Substituting $R_c$ back into Eq.~(\ref{eq:edep}) gives the threshold energy,
\begin{equation}
    E_{\rm th}(T) = \frac{\pi \beta T^4}{3}\,R_c^3(T)
    = \frac{\pi \beta T^4}{3}
    \left[\frac{\alpha\,\tau_0\,\beta\,T^4}{4\,\rho_s\,\Delta E_{\rm Zee}}\right]^{3/2}
    \exp\!\left(\frac{3\tilde{U}(B)}{2k_B T}\right).
\label{eq:eth_of_T}
\end{equation}
This equation can then be mimimized over $T$ to find the lowest temperature $T^*$ that satisfies the feedback condition. This yields a formula for the threshold as a function of $B$,
\begin{equation}
E_{\rm th}(T^*,B) \sim 10^{-5}\, \beta^{5/2}   \left[\frac{\alpha\,\tau_0}{\rho_s\,\Delta E_{\rm Zee}(B)}\right]^{3/2}   \left[\frac{\tilde{U}(B)}{k_B}\right]^{10}  \,.
\end{equation}

The $\mathcal{O}(1)$ prefactor in the ignition condition Eq.~(\ref{eq:ignition}) is not captured by this analytic estimate. The three-dimensional simulations of Sec.~3 serve precisely to calibrate this prefactor: by identifying the empirical threshold in the simulations and comparing to Eq.~(\ref{eq:eth_of_T}), we establish how accurately the analytic expression predicts the true onset of runaway behaviour.

\section{Simulation Tests of the Stochastic Ignition Condition}

The analytic framework of Sec.~2 predicts a  threshold energy given by Eq.~(\ref{eq:eth_of_T}), below which an avalanche cannot self-sustain and above which positive feedback drives runaway.  To validate the analytic treatment above against a concrete dynamical model, we perform three-dimensional simulations of coupled thermal diffusion and stochastic spin relaxation.

\subsection{Numerical Framework}

We simulate the evolution of the temperature field and individual spins in a cubic periodic volume of side length $L$, discretized with $N_g^3$ grid cells. The temperature is defined on the grid as $T(x,y,z)$ with spacing $dx_g = L/N_g$. Spins are located on a lattice with unit-cell spacing $R_s$, each initially prepared anti-aligned with the external magnetic field $B$. The background temperature is uniform and set to $T_i$.

The simulation is initiated by depositing an energy $E_i$ within a single unit cell at a random location in the box, producing a local temperature increase
\begin{equation}
\delta T_i = \left(\frac{E_i}{R_s^3\, c_s}\right)^{1/4},
\end{equation}
where $c_s$ is the specific heat per unit cell.

Thermal evolution is governed by the heat diffusion equation, which we write in terms of the variable $u \equiv T^4$. For a phonon-dominated specific heat $c_v \propto T^3$ and a temperature-independent thermal diffusivity $\alpha$, the heat equation becomes linear in $u$:
\begin{equation}
\partial_t T^4(\mathbf{x},t) = \alpha \nabla^2 T^4(\mathbf{x},t),
\end{equation}
allowing an efficient spectral (FFT-based) time evolution. The update over a timestep $\delta t$ is:
\begin{equation}
T^4(t+\delta t) = \mathcal{F}^{-1}\!\left[e^{-\alpha \mathbf{k}^2 \delta t}\,\mathcal{F}[T^4(t)]\right],
\end{equation}
where $\mathcal{F}$ denotes the discrete Fourier transform. If $\alpha$ depends significantly on temperature, or if the heat capacity deviates from $T^3$ scaling, the evolution equation becomes nonlinear and a more general numerical treatment would be required; we leave such extensions to future work, noting that the measurements of Sec.~4 inform where and how strongly these deviations occur in practice.

After each diffusion step, every spin that has not yet flipped is assigned a probability to relax during the timestep $\delta t$:
\begin{equation}
P_i = \frac{\delta t}{\tau_0 \exp\!\left(\dfrac{\tilde{U}(B)}{k_B T(\mathbf{r}_i)}\right)},
\end{equation}
where $T(\mathbf{r}_i)$ is the temperature interpolated at the position of the $i$th spin. For each spin, a uniform random variable $U_i \in [0,1]$ is drawn; if $U_i < P_i$, the spin is flipped. Each spin flip releases Zeeman energy $2J\mu_B B$, deposited locally as a temperature increment
\begin{equation}
\delta T_f = \left(\frac{2J\mu_B B}{R_s^3\,c_s}\right)^{1/4}
\end{equation}
within the corresponding unit cell. A single timestep therefore consists of: (i)~thermal diffusion, (ii)~assignment of spin-flip probabilities, (iii)~stochastic spin updates, and (iv)~local heat injection from flipped spins. The timestep $\delta t$ is chosen to satisfy $\delta t \ll \tau_0$ and $\delta t \ll L^2 /(N_g^2\alpha)$ to ensure numerical stability and resolution of the relaxation dynamics.

\subsection{Validation of the Refined Ignition Criterion}

\begin{figure*}[!ht]
    \includegraphics[width=0.97\textwidth]{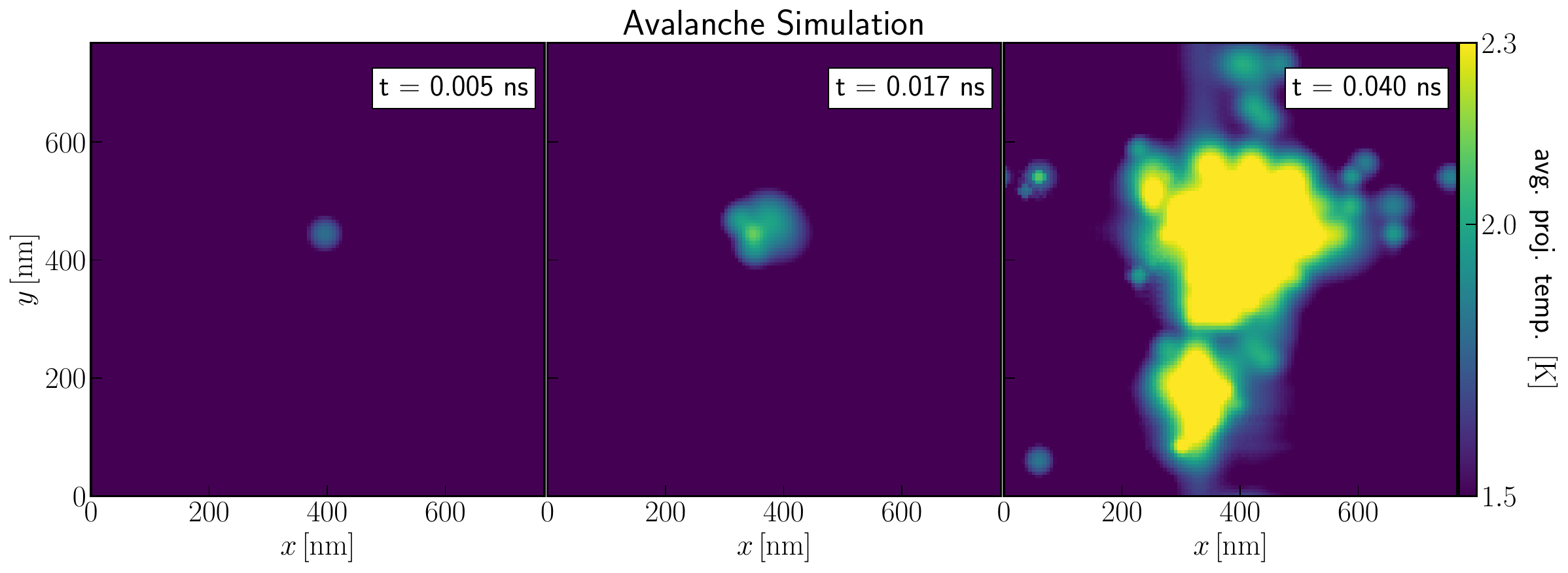}
    \caption{Simulation of a stochastic avalanche. Each panel shows the average projected temperature field at a different time $t$. \textit{Left}: the initial deposited energy spreads diffusively. \textit{Centre}: the temperature rise causes a small fraction of nearby spins to flip stochastically, releasing additional Zeeman energy. \textit{Right}: the resulting positive feedback drives a runaway avalanche front. Simulation parameters: grid resolution $N_g = 128^3$, unit cell size $R_s = 6\,\mathrm{nm}$, thermal diffusivity $\alpha = 10^{-6}\,\mathrm{m^2\,s^{-1}}$, magnetic field $B = 6.7\,\mathrm{T}$, initial temperature $T_i = 0.1\,\mathrm{K}$, attempt time $\tau_0 = 0.1\,\mathrm{ns}$, deposited energy $E_i = 0.01\,\mathrm{eV}$, spin $J = 21/2$, and specific heat coefficient $\beta = 10^{-9}\,\mathrm{eV\,K^{-4}\,nm^{-3}}$.}
    \label{fig:avalanch}
\end{figure*}

Figure~\ref{fig:avalanch} shows representative snapshots of the temperature evolution in a simulation where, at the peak temperature following the initial deposition, the timescale hierarchy satisfies $t_d < t_f$. Under the conservative ``hard'' criterion of earlier estimates, this regime would not be expected to produce an avalanche.

Nevertheless, the simulation exhibits clear runaway behavior. The initially deposited energy diffuses outward, but during the diffusion window a small fraction of spins relaxes stochastically, releasing Zeeman energy. This additional heating locally elevates the temperature, which in turn shortens the relaxation time and increases the fraction of spins that flip. The positive feedback between stochastic spin relaxation and thermal diffusion leads to the formation of a propagating avalanche front.

The onset of runaway behavior coincides with the cumulative released Zeeman energy becoming comparable to the energy required to sustain the local temperature against diffusion losses, in quantitative agreement with the analytic ignition condition of Eq.~(\ref{eq:ignition}). In particular, ignition occurs even though the strict ordering $t_f < t_d$ is not satisfied, confirming that avalanche formation is controlled by the integrated energy release during the diffusion window rather than by a hard comparison of timescales.

These simulations therefore validate the central theoretical result of this work: stochastic spin relaxation parametrically lowers the effective energy threshold for avalanche formation relative to the conservative hard criterion of Ref.~\cite{Bunting:2017}. The simulation parameters used here are representative of the Mn$_{32}$ material characterised in Sec.~4. The simulated threshold of $\sim 0.01$~eV is consistent with the analytic estimate $E_{\rm th}(T)$ from Eq.~(\ref{eq:eth_of_T}), evaluated using the measured values of $\tau_0$, $\beta$, and $\tilde{U}(B)$ for that material.

\section{Experimental Characterisation of Low-Temperature Thermal Properties of SMMs}
\label{sec:materials}

The stochastic avalanche framework developed in Secs.~2 and~3 takes as direct inputs the specific heat $c(T)$ and the thermal diffusivity $\alpha(T)$ at cryogenic temperatures. The analytic treatment of Sec.~2 assumes that, at sufficiently low temperatures, the dominant contribution to the specific heat is the Debye lattice term,
\begin{equation}
c_{\rm ph}(T) \propto T^3.
\end{equation}
This is precisely the regime in which the avalanche model is self-consistent: the $T^3$ dependence of $c_v$ gives rise to the linear heat equation in $T^4$ exploited by the spectral integrator of Sec.~3. Deviations from this scaling---whether from hyperfine contributions, Schottky-type magnetic terms, or anomalous phonon scattering---would modify both the analytic threshold estimate and the numerical evolution. The measurements reported in this section directly characterise these deviations and bound their impact on the inferred ignition thresholds.

The measurements were carried out using cryogenic calorimetry and thermal-transport techniques optimized for small SMM single crystals. Specific heat was measured on millimeter-scale crystals using a long relaxation calorimetry method~\cite{tanaka:2022,Mizukami2023} in $^3$He and dilution-refrigerator cryostats, enabling measurements down to $\sim 0.2~\mathrm{K}$ with microgram-scale samples. Thermal diffusivity was characterized via micro-temperature wave analysis using a single-crystal sample approximately 100$\mu$m in size, while the sample was cooled down to 9 K with a Gifford–McMahon (GM) refrigerator~\cite{10.1063/5.0055707,doi:10.1021/jacs.3c07921,doi:10.1021/jacs.4c11849,doi:10.1021/acs.jpclett.5c00354}. These measurements directly determine $\alpha(T)$, which enters the diffusion time $t_d \sim R^2/\alpha$ in the critical radius expression Eq.~(\ref{eq:rc}).

These measurements provide the first material benchmarks in the broader program aimed at engineering SMM crystals with optimised relaxation and thermal properties, with the ultimate goal of approaching meV-scale detection thresholds.

\subsection{Mn$_{12}$ac: Heat Capacity and Thermal Diffusivity}

We first examine single-crystalline Mn$_{12}$ac, a well-characterised SMM that serves as a benchmark system. The spin relaxation follows an Arrhenius form
\begin{equation}
t_f(T) = \tau_0 \exp\!\left(\frac{\tilde{U}(B)}{k_B T}\right).
\end{equation}
From AC magnetization measurements we obtain
\begin{equation}
\tilde{U}(B=0) = 68.1~\mathrm{K},
\qquad
\tau_0 \simeq 1.2 \times 10^{-7}~\mathrm{s},
\end{equation}
consistent with previously reported values~\cite{novak:1995}. We decompose the low-temperature specific heat as
\begin{equation}
c(T) = c_{\rm ph} + c_n + c_{\rm spin},
\end{equation}
where $c_{\rm ph} \propto T^3$ is the Debye phonon contribution, $c_n \propto T^{-2}$ is the nuclear hyperfine contribution, and $c_{\rm spin}$ is approximated by a two-level Schottky model,
\begin{equation}
c_{\rm spin} \propto
\left(\frac{\Delta}{T}\right)^2
\frac{e^{-\Delta/T}}{\left(1 + e^{-\Delta/T}\right)^2}.
\end{equation}

\begin{figure}[t]
\centering
\includegraphics[width=0.7\linewidth]{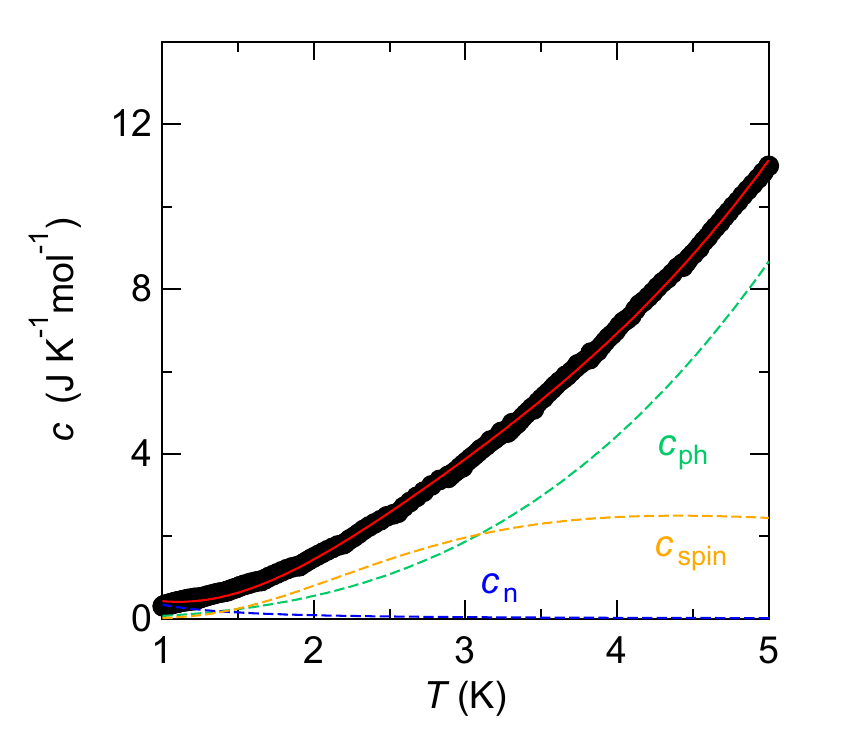}
\caption{Temperature dependence of the specific heat of Mn$_{12}$ac (points), together with the decomposition into Debye phonon ($c_{\rm ph} \propto T^3$), nuclear hyperfine ($c_n \propto T^{-2}$), and Schottky spin contributions (lines). The low-temperature $T^3$ regime validates the Debye assumption underlying the avalanche model of Sec.~2.}
\label{fig:thermal_props}
\end{figure}

\begin{figure}[t]
\centering
\includegraphics[width=0.7\linewidth]{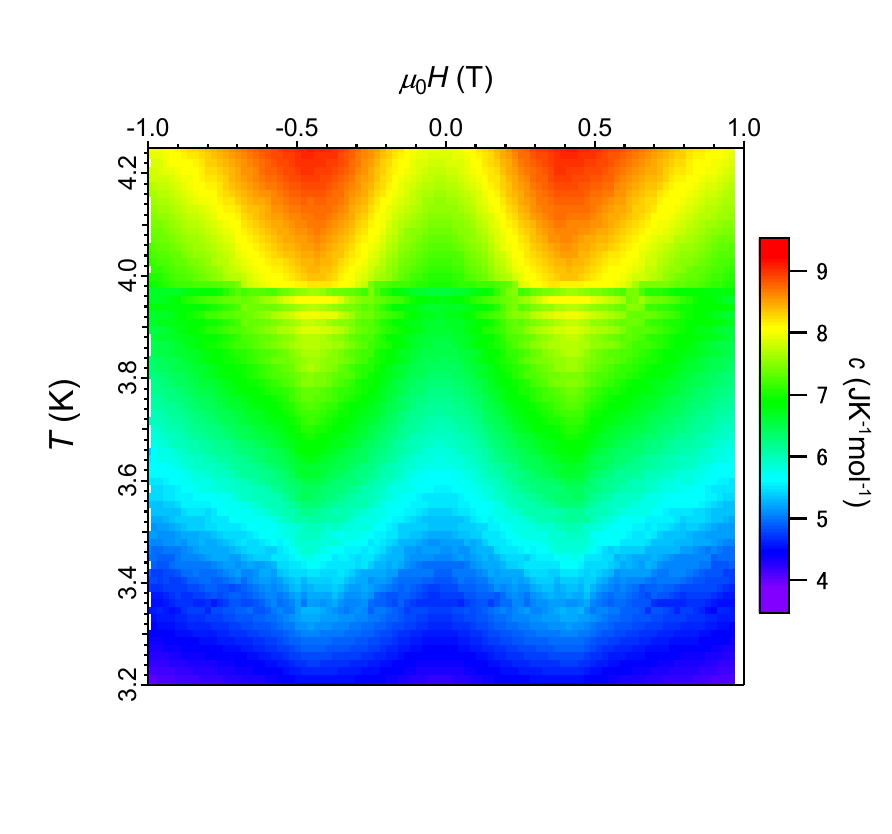}
\caption{Specific heat of Mn$_{12}$ac as a function of applied magnetic field and temperature. Field-dependent peaks correspond to resonant quantum tunneling between spin levels.}
\label{fig:mn12_tunnel}
\end{figure}

The measured temperature dependence is shown in Fig.~\ref{fig:thermal_props}. Fitting the decomposition above yields a spin excitation scale $\Delta = 10.6~\mathrm{K}$. At the lowest temperatures the data exhibit the expected Debye-like lattice contribution $c_{\rm ph} \propto T^3$, while deviations at intermediate temperatures are well-captured by the combined hyperfine ($\propto T^{-2}$) and Schottky-like spin contributions. In the detector operating regime $T \ll \Delta$, the Schottky contribution is exponentially suppressed and the phonon term dominates, validating the $c(T) \propto T^3$ assumption used throughout Secs.~2 and~3.

An additional feature visible in the data is the appearance of magnetic field-dependent peaks in the specific heat, displayed in Fig.~\ref{fig:mn12_tunnel}. These peaks occur at resonant fields and correspond to quantum tunneling between spin levels. While tunneling is not required for avalanche triggering in our model, its thermodynamic signature confirms the sensitivity of the calorimetry to spin-level structure and provides an independent probe of relaxation channels.

\subsubsection*{Thermal diffusivity of Mn$_{12}$ac}

\begin{figure}[t]
\centering
\includegraphics[width=0.5\linewidth]{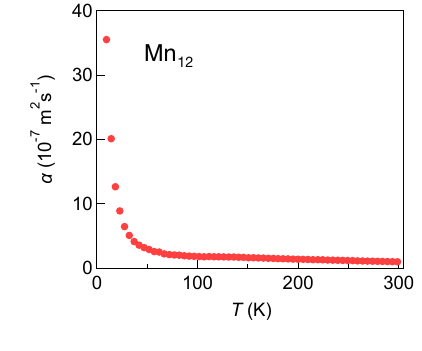}
\caption{Thermal diffusivity $\alpha(T)$ of Mn$_{12}$ac single crystals.}
\label{fig:mn12alpha}
\end{figure}

The measured thermal diffusivity  of Mn$_{12}$ac exhibits a nontrivial temperature dependence over the experimentally accessed range, whereas we assumed a constant-$\alpha$ in the simulations of Sec.~3. That choice was made to enable a linear evolution equation in $T^4$ and the use of an efficient FFT-based solver. 
The present measurements support the numerical choice of constant-$\alpha$ used in the simulation of the threshold; the value of $\alpha$ used in the Mn$_{32}$ simulations was chosen to be representative of the range measured here for Mn$_{12}$ac, as direct measurements for Mn$_{32}$ are not yet available.  

We note that temperature of these measurements did not reach the sub-Kelvin regime most relevant for detector operation, where the detailed behaviour of $\alpha(T)$ enters the critical radius $R_c(T)$ of Eq.~(\ref{eq:rc}). The observed rise of $\alpha$ as temperature decreases reflects the growth of the phonon mean free path, $\ell$. However, once $\ell$ saturates at the boundary- or defect-limited scale, we expect that $\alpha$ would become approximately temperature-independent, and thus expect the constant-$\alpha$ idealization adopted in Sec. 3 for the sub-Kelvin operating regime is well-justified. Characterising $\alpha(T)$ in this regime --- for both Mn$_{12}$ac and Mn$_{32}$ --- and including its detailed temperature dependence in numerical simulations is a priority for the next stage of this program.

\subsection{Mn$_{32}$: Fast Spin Relaxation and Sensitivity Prospects}

Mn$_{12}$ac serves as a useful benchmark system; however, its relatively large attempt time $\tau_0 \simeq 1.2 \times 10^{-7}\,\mathrm{s}$ implies relaxation times $t_f$ that remain parametrically longer than the characteristic diffusion times $t_d$ relevant for nanometer-scale heated regions. In this regime $t_f \gg t_d$, and although the stochastic mechanism described in Sec.~2 lowers the ignition threshold relative to the hard criterion, the reduction is not sufficient to approach the meV energy scale required for sub-GeV dark matter sensitivity.

By contrast, Mn$_{32}$ provides a qualitatively different operating regime. Its substantially smaller attempt time, $\tau_0 \simeq 1.5 \times 10^{-11}\,\mathrm{s}$, moves the intrinsic relaxation dynamics much closer to the diffusion window, with $t_f$ approaching $t_d \sim 10^{-10}\,\mathrm{s}$ for meV energy deposits. This proximity enhances the stochastic relaxing fraction $f$ during the diffusion interval and therefore increases the integrated Zeeman energy available to seed runaway behavior. As a result, Mn$_{32}$ emerges as a promising candidate material capable of significantly lowering the avalanche threshold beyond the MeV-scale demonstrations of Refs.~\cite{Chen:2020,Kohn:2025}, and represents a concrete  target for substantially improved detector sensitivity.

In this section we measure the low-temperature heat capacity of Mn$_{32}$, providing the material inputs required for the threshold expression Eq.~(\ref{eq:eth_of_T}). Using the experimentally determined $c_v(T)$ together with the measured relaxation parameters, we evaluate the corresponding avalanche threshold and assess the projected dark matter sensitivity of this material under realistic operating conditions.

AC magnetization measurements yield
\begin{equation}
\tilde{U}(B=0) = 44.6~\mathrm{K},
\qquad
\tau_0 \simeq 1.5 \times 10^{-11}~\mathrm{s},
\end{equation}
where $\tilde{U}(B=0)$ is consistent with the previously reported value~\cite{manoli:2011}. The attempt time is four orders of magnitude smaller than in Mn$_{12}$ac, placing Mn$_{32}$ in a fundamentally different regime for the avalanche threshold.

\begin{figure}[t]
\centering
\includegraphics[width=0.6\linewidth]{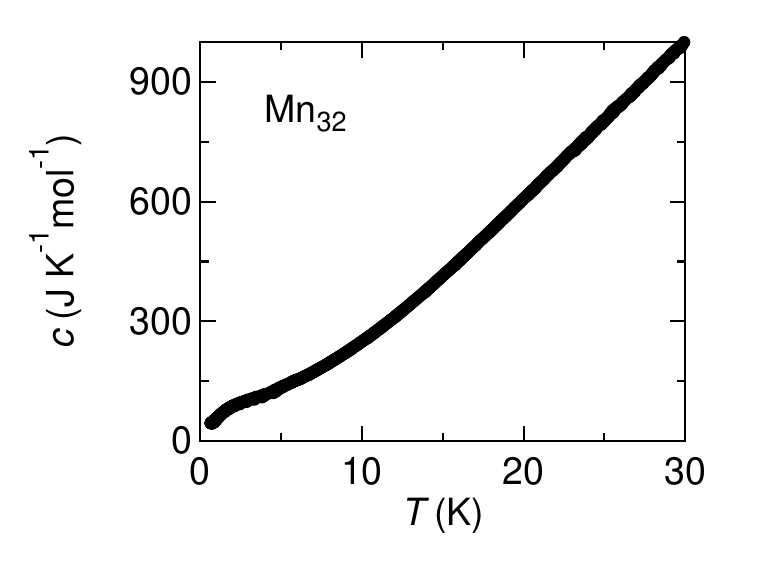}
\caption{Specific heat of Mn$_{32}$ single crystals as a function of temperature. The pronounced low-temperature hump is consistent with a Schottky-like contribution from low-lying spin levels (gap $\Delta \approx 5.8\,\mathrm{K}$). The large magnitude of $c(T)$ near $1\,\mathrm{K}$ sets the energy scale for local temperature excursions and enters directly into the avalanche threshold of Eq.~(\ref{eq:eth_of_T}).}
\label{fig:mn32_c}
\end{figure}

The specific heat of Mn$_{32}$ is shown in Fig.~\ref{fig:mn32_c}. A pronounced low-temperature hump is visible, consistent with a Schottky-like contribution, from which we extract a gap
\begin{equation}
\Delta \approx 5.8~\mathrm{K}.
\end{equation}

However, Eq. (19) does not provide a satisfactory fit to the Mn$_{32}$ data, and the origin of this discrepancy remains unclear. One possible reason is that the spin-level splitting induced by the axial anisotropy term is smaller than that in Mn$_{12}$, in which case the spin-specific-heat contribution may not be well approximated by the two-level Schottky model of Eq. (20).

The magnitude of the heat capacity near $1\,\mathrm{K}$ -- the typical order of magnitude of temperature achieved in an avalanche -- is substantially larger than in Mn$_{12}$ac, which is consistent with the previous report~\cite{PhysRevB.79.104414}. 
The Schottky contribution is again suppressed for $T \ll \Delta$; this can be checked consistently in simulations that include only the phonon $T^3$ dependence. We note, however, that when the SMM crystal is being used as a detector, it is initially in a metastable state, and so the relevant heat capacity used in the avalanche threshold equations is not the same as the one measured in equilibrium. That is, after an initial local energy deposit that triggers an avalanche, the system is far from equilibrium. This suggests that even if the local temperature rises to be comparable to or greater than $\Delta$ during an avalanche, the Schottky term would not be fully active. We postpone an estimation of its effect---expected to be small---to a future more detailed numerical simulation.

Using these experimentally measured relaxation and thermal parameters as inputs to the stochastic avalanche framework, we find that avalanches can be triggered by initial energy deposits as small as $\sim 0.01\,\mathrm{eV}$ under representative operating conditions. This demonstrates that materials with intrinsically fast relaxation dynamics can substantially reduce the effective ignition threshold relative to previously studied systems, and points toward a route to sub-eV sensitivity in future engineered SMM detectors.

\section{Conclusion}

We have presented a refined theoretical and experimental framework for assessing the sensitivity of SMM-based dark matter detectors. The central theoretical result is that the effective ignition threshold for a magnetic avalanche is significantly lower than estimated by the conservative hard criterion of Ref.~\cite{Bunting:2017}. The main point is that spin relaxation is stochastic: even when the mean relaxation time $t_f$ exceeds the thermal diffusion time $t_d$, a fraction $f \approx t_d/t_f$ of spins relax during the diffusion window, releasing Zeeman energy that scales as $R^5$. Setting this released energy equal to the deposited energy defines a critical radius $R_c$ (Eq.~\ref{eq:rc}), which when substituted back yields a closed, explicit expression for the energy threshold $E_{\rm th}(T)$ (Eq.~\ref{eq:eth_of_T}) in terms of directly measurable material parameters. 

The 3D coupled thermal--spin simulations of Sec.~3 validate this analytic framework. On the experimental side, low-temperature calorimetric measurements of Mn$_{12}$ac confirm the Debye-like $c(T) \propto T^3$ behaviour assumed in the analytic treatment and required for the spectral simulation method. Thermal diffusivity measurements ground the choice of $\alpha$ used in the simulations and motivate extending the characterisation to sub-Kelvin temperatures, where $\alpha(T)$ most directly controls the threshold via $R_c(T)$.

The characterisation of Mn$_{32}$---with an attempt time $\tau_0$ four orders of magnitude shorter than Mn$_{12}$ac---yields simulated avalanche thresholds at the $\mathcal{O}(10^{-2})\,\mathrm{eV}$ scale. This illustrates a concrete materials path toward sub-eV sensitivity. More broadly, these results establish the analytic and experimental tools needed to systematically screen and engineer SMM candidates for optimised dark matter detection performance.

\section*{Methods}
The specific heat measurements were performed using the long-time relaxation method described in the text. The uncertainty associated with the data acquisition itself is negligibly small for each individual data point. On the other hand, when converting the measured heat capacity into specific heat capacity, the molar amount estimated from the sample mass is used, and the heat capacity is divided by this molar amount. Consequently, a systematic multiplicative error is introduced over the entire dataset due to the uncertainty in the sample mass measurement. This error is estimated to be on the order of a few percent, up to approximately 5 percent. Such a systematic error does not affect the analysis of the temperature dependence of the specific heat discussed in the main text.

\section*{Acknowledgments}

AE is supported by the Center for Data-Driven Discove seed grant and a is a Kavli Fellow. This work was supported by Grants-in-Aid for Scientific Research (KAKENHI) (JP22K18712, JP23K26522, JP26H02230, 26K17135) from Japan Society for the Promotion of Science (JSPS), and FOREST (JPMJFR236O) from Japan Science and Technology (JST), and by Japan Science and Technology Agency (JST) CREST (JPMJCR18I4 for Ta.F.). This work is supported by World Premier International Research Center
Initiative (WPI Initiative), MEXT, Japan. This work was supported by the U.S.~Department of Energy~(DOE), Office of Science, National Quantum Information Science Research Centers, Superconducting Quantum Materials and Systems Center~(SQMS) under Contract No.~DE-AC02-07CH11359.  S.R.~is supported in part by the U.S.~National Science Foundation~(NSF) under Grant No.~PHY-1818899.
S.R.~is also supported by the Simons Investigator Grant No.~827042.

\end{document}